# Molecular Gas Adsorption Induced Carrier Transport Studies of Epitaxial Graphene using IR Reflection Spectroscopy

B.K. Daas[1a], W.K. Nomani[1], K.M. Daniels[1], T.S. Sudarshan[1], Goutam Koley[1] and MVS Chandrashekhar[1]

[1]Department of Electrical Engineering, University of South Carolina, Columbia SC 29208, USA

Daas@email.sc.edu[a]

**Abstract:** We investigate molecular adsorption doping by electron withdrawing $NO_2$ and electron donating $NH_3$ on epitaxial graphene grown on C-face SiC substrates. Amperometric measurements show conductance changes upon introduction of molecular adsorbents on epitaxial graphene. Conductance changes are a trade-off between carrier concentration and scattering, and manifest at direct current and optical frequencies. We therefore investigate changes in the infrared (IR) reflection spectra to correlate these two frequency domains, as reflectance changes are due to a change of epitaxial graphene (EG) surface conductance. We match theory with experimental IR data and extract changes in carrier concentration and scattering due to gas adsorption. Finally, we separate the intraband and interband scattering contributions to the electronic transport under gas adsorption. The results indicate that, under gas adsorption, the influence of interband scattering cannot be neglected, even at DC.

**Introduction**

Graphene, a two-dimensional (2D) honeycomb carbon crystal, the basic building block of other $sp^2$ carbon nanomaterials, such as nanographite sheets and carbon nanotubes, exhibits unusual electronic and optical properties [1-3]. Graphene has a unique electronic band structure, with the conduction band touching the valance band at the K and K' points in the Brillouin zone. This gives graphene its characteristic linear Dirac-Fermion energy dispersion (E=hk$v_F$), where E is the energy, h is Planck's constant, k is the wave vector and $v_F$ (~$10^6$m/s) is the Fermi velocity. Therefore electrons in the ideal graphene sheet exhibit very high electron mobility [4]. This truly 2D structure with high tunable semi-metallic conductivity have made graphene ideal candidate for gaseous and vapor sensing applications. The sensitivity of exfoliated graphene to $NH_3$ and $NO_2$ [5] and that of epitaxial graphene [6] have been established by molecular adsorption doping. Due to adsorption of impurities on epitaxial graphene (EG) surface, the electron concentration can either increase or decrease depending on whether the adsorbent is electron donating or withdrawing. These adsorbents can also influence carrier scattering, all of which change the surface conductivity. In our earlier paper [7] we established a change of conductivity due to molecular adsorption doping by DC measurements. In this paper we emphasize carrier transport studies through Fourier transform infrared (FTIR) measurement for molecular adsorption doping. Theoretically, we account for the influence of surface impurities on the conductivity of EG by molecular gas adsorption [5] using the Boltzmann picture. In our previous work [7] we have shown evidence of a surface plasmon polariton at the EG/SiC interface through IR reflection spectra. Here we investigate the carrier transport by FTIR reflection spectroscopy along with a mathematical model to extract various transport parameters, including the influence of interband scattering, which is non-negligible at IR frequencies and high surface impurity concentrations.

**Experiment:**

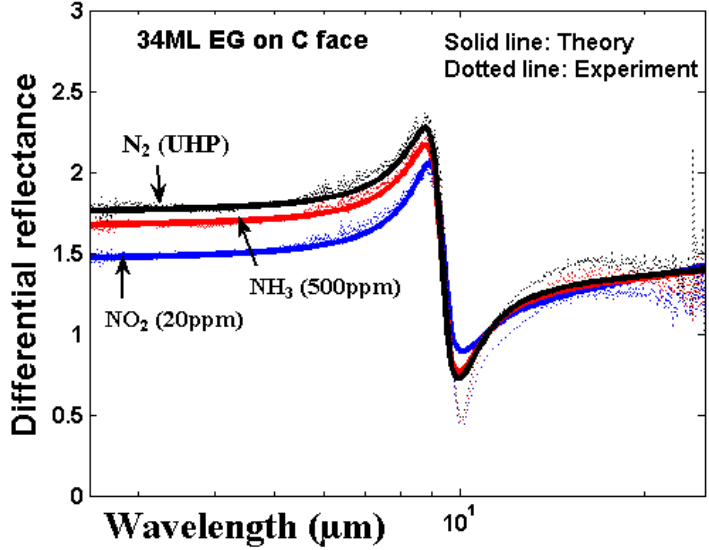

**Figure-1:** IR reflection measurement on 34 ML EG on C face in $N_2$, $NH_3$ and $NO_2$ environments (SiC reference).

Epitaxial growth of large-area graphene by thermal decomposition of commercial <0001> 4H and 6H SiC substrates at high temperature and vacuum has been demonstrated [8]. This produces EG a few monolayer, ML to >50 ML thick, depending on growth conditions. In our experiments, EG was grown on commercial $n^+$ $8^0$ off axis 4H-SiC substrates on C-face, nitrogen doped ~$10^{19}$/cm$^3$. 1cmx1cm samples were degreased using trichloroethylene, acetone and methanol respectively. They were then rinsed in DI water for three minutes. The samples were finally dipped in HF for two minutes to remove native oxide and rinsed with DI water before being blown dry. They were then set in the crucible in an inductively heated furnace where high vacuum was maintained (<$10^{-6}$ Torr) and baked out at $1000^0$C for 13 to 15 hours. The temperature was slowly raised to the growth temperature (1250-1400°C). All growths were performed for 60 minutes before cooling to $1000^0$C at a ramp rate of 7~$8^0$C/min and eventually to room temperature. Slow temperature ramps were utilized to minimize thermal stress on the samples.

After growth AFM (atomic force microscopy) was used to investigate the EG surface morphology with and the morphology is similar to that observed by others [4,7-8]. Raman measurements were carried out on EG on both carbon (C) and silicon (Si) faces. Micro-Raman spectroscopy using a 632nm laser shows the G peak (~1590cm$^{-1}$), D peak (~1350cm$^{-1}$) and 2D peak (~2700cm$^{-1}$) characteristic of EG [9]. The ratio of intensities of the D-peak to G-peak, $I_D/I_G \leq 0.2$ demonstrates the high quality of our graphene [9]. X-ray photoelectron spectroscopy (XPS) measurements were done to obtain the thickness in monolayer's (ML) on EG. The thickness extracted by XPS was consistent with our FTIR measurements [10].

**Results and Discussion:**

$NO_2$ is a strong oxidizer with electron withdrawing capability, and is expected to decrease electron carrier concentration on the EG surface, while the converse is true for $NH_3$ as it is electron donating. However, our measurements are based on EG grown on C-face SiC, which is 34 ML thick, much greater than the 1ML screening length in graphene. Thus, this layer is expected to be close to electrically neutral i.e. the Fermi level is close to the Dirac point (K and K' in the Brillouin zone). Therefore, carrier concentration will increase in both the cases regardless of whether the adsorbate is electron withdrawing or electron donating. Polar molecules [6] change EG conductivity by a) inducing carriers in the EG and b) increasing scattering i.e. decreasing mobility. Surface carrier concentration in EG is given by

$$n_s = \int_0^\infty D(E) f(E - E_F) dE \qquad (1)$$

where D(E) is the density of states and Fermi energy function is $f(E-E_F)$, $E_F$ is the Fermi level. Scattering ($\tau$) includes both intra and interband scattering [11]. Increase or decrease of conductivity is a tradeoff between carrier concentration and scattering. Thus, it is critical to investigate the role of $E_F$ and $\tau$ in response to gas adsorption.

Fourier transform infrared reflection (FTIR) measurements were carried out with a blank SiC substrate, cut from the same wafer as the grown samples, as the reference. All reflectance measurement presented here are the differential reflectance with respect to the SiC substrate. ML, $E_F$ and $\tau$ were extracted by fitting measurement to theory [7]. A more detailed analysis of this data allowed the extraction of the influence of interband scattering as well, which is non-negligible at IR frequencies and high surface impurity concentrations. Fig-1 shows the IR reflection spectra for ~34ML EG in $N_2$, $NH_3$ and $NO_2$ environment where we find that reflectance (indicative of conductivity, with higher conductivity leading to higher reflectance) changes for different gases due to adsorption of surface impurities. Reflectivity decreases both in $NO_2$ and $NH_3$ environments compared to $N_2$ with $NO_2$ showing greater decrease than $NH_3$. We extract ML, $E_F$ and intra and inter band scattering time by matching experimental data with the theory. We also modified our earlier proposed mathematical model [7] to account for impurity adsorption where we match optical conductivity [12] at high frequency with conductivity in the Random Phase approximation (RPA) to extract surface impurity concentration, $n_i$ by

$$\sigma_{T=0}^{RPA} = \frac{e^2}{\pi h}[\frac{n_s}{n_i G[4r_s/(2-\pi r_s)]} + \frac{n_i F[4r_s/(2-\pi r_s)]}{4n_s}] \qquad (2)$$

where h is Planck's constant,

$$r_s = \frac{e^2}{4\pi\varepsilon_0\varepsilon_{SiC}v_F\hbar} \qquad (3)$$

where e is the electron charge $1.6\times10^{-19}C$, $v_F$ is the Fermi velocity $1.1\times10^6$ m/s, $\varepsilon_{SiC}$ is the dielectric constant of SiC which has different values for high frequency ($\varepsilon_{SiC}$~6.5) and low frequency ($\varepsilon_{SiC}$~9.52) regime and G and F function defined [12] as,

$$G(x) = \frac{x^2}{8}\int_0^{2\pi}\frac{\sin^2\theta}{(\sin\frac{\theta}{2}+x)^2}d\theta \quad \text{and} \quad F(x) = \frac{x^2}{8}\int_0^{2\pi}\frac{(1-\cos\theta)^2}{(\sin\frac{\theta}{2}+x)^2}d\theta \qquad (4)$$

Considering two limiting values of SiC dielectric constant (high frequency~6.5 and low frequency~9.52), two different values of $r_s$ (high frequency~0.31 and low frequency~0.21) were calculated. For the high frequency $r_s$=0.31 was used for the conductivity

$$\sigma_{T=0}^{RPA}[high frequency] = \frac{e^2}{\pi h}[\frac{n_s}{n_i G[4r_s/(2-\pi r_s)]}] \qquad (5)$$

matching with the optical conductivity to extract $n_i$. A similar procedure was used for the low frequency side where $r_s$~0.21 arises from $\varepsilon_{SiC}$~9.52 while considering

$$\sigma_{T=0}^{RPA}[low frequency] = \frac{e^2}{\pi h}[\frac{n_i F[4r_s/(2-\pi r_s)]}{4n_s}] \qquad (6)$$

**Table-1: Our extracted parameters indicate value of Fermi level, impurity and scattering in three different gas environment when IR reflection experiment data matches with theory.**

| Gas | Fermi level (meV) | Impurity ($cm^{-2}$) | Intraband scattering time (s) | Interband scattering time(s) |
|---|---|---|---|---|
| $N_2$ | 25 | $2 \times 10^{11}$ | 2.8 to $9 \times 10^{-14}$ | 6 to $2.7 \times 10^{-14}$ |
| $NH_3$ | 30 | $6 \times 10^{12}$ | 9e-15 to $6 \times 10^{-15}$ | 2 to $1.6 \times 10^{-15}$ |
| $NO_2$ | 35 | $2 \times 10^{13}$ | 2e-15 to $9 \times 10^{-16}$ | 5 to $2.6 \times 10^{-16}$ |

$n_i$ extracted at both these frequency regimes was consistent. For further confirmation, we calculate intra and inter band scattering from $n_i$ using equations presented elsewhere [12] and was found to be consistent with our extracted data within the experimental error limit.

Table 1 show the extracted carrier transport parameters as a function of gas adsorbents. With $N_2$ gas, Fermi level is ~25meV, close to neutral because our EG is thick (34ML). As $N_2$ is an inert gas and should not contribute any impurity on the EG surface. Our extracted parameters indicates a surface impurity concentration of $2x10^{11}cm^{-2}$, fairly consistent with an ex-situ sample that has not had any degassing or other processing performed on it.

For $NH_3$ and $NO_2$, the surface impurity concentration is higher than $N_2$ because of the nature of the gas interaction with the carriers on the EG surface. In both cases, intra and interband scattering time shortens, due to more scattering from the adsorbed impurity. Overall conductivity decreases both for $NH_3$ and $NO_2$ compared to $N_2$ environment. With $NO_2$, overall conductivity decreases more than $NH_3$ despite a strong increase in carrier concentration, indicating that the EG surface has greater affinity for $NO_2$ adsorption, leading to much greater carrier scattering at the adsorption surface.

In summary, we find that the infrared carrier transport in EG can be described only by accounting for the influence of interband scattering. In doing so, we find a self-consistent explanation of carrier transport in EG, where carrier scattering (interband and intraband) increases in response to surface impurities, along with a change in carrier concentration caused by charge transfer between the molecules and EG. This leads to an increase/decrease in overall conductivity/reflectivity depending on the tradeoff between scattering and carrier concentration

**Acknowledgement:**
The authors would like to acknowledge the Southeastern Center for Electrical Engineering Education, and the National Science Foundation (NSF) ECCS-EPDT Grant #1029346 under the supervision of program director Rajinder Khosla for funding this work.